\documentclass[preprint,epsfig,aps,showpacs]{revtex4}
\usepackage{amsmath}
\usepackage{graphicx}

\newcommand{\f}{\begin{equation}}
\newcommand{\ff}{\end{equation}}
\newcommand{\fa}{\begin{eqnarray}}
\newcommand{\ffa}{\end{eqnarray}}

\begin{document}
\title{Big Rip in ${\bf SO(1,1)}$ phantom universe}
\author{Yi-Huan Wei${ }^{1}$}
\affiliation{%
${ }^1$ Department of Physics, Bohai University, Jinzhou 121000,
 Liaoning, China}

\begin{abstract}
~~For the inverse linear potential, the $SO(1,1)$ field behaves as
phantom for late time and the Big Rip will occur. The field
approaches zero as time approaches the Big Rip, here. For this
potential the phantom equation of state takes the late-time
minimum $w_\Phi=-3$. We give some discussions that the Big Rip in
the $SO(1,1)$ model may be treated as either the transition point
of universe from expansion to extract phase or the final state. In
the latter picture of the universe, the field has the $T$ symmetry
and the scale factor possesses the $CT$ symmetry, for which the
$SO(1,1)$ charge $\bar{Q}$ plays a crucial role.

\end{abstract}

\maketitle

\section {Introduction}

The recent type Ia supernova (SN Ia) observations \cite{Riess} and
the cosmic microwave background radiation (CMBR) measurements
\cite{Ber} indicates that the universe is expanding acceleratedly
and thus there exists the dark energy in it
\cite{Rathra,Turner,Boyle,Bento,Caldwell,Caldwel,McInnes}. Dark
energy may be very likely to possess a super-negative pressure,
i.e., the phantom energy
\cite{Caldwell,Caldwel,McInnes,P,Nojiri,SNojiri,Odintsov,GuoZK,A,Ca,Townsend,Onemli,Carvalho,Gonz,Li}.
Phantom violates the weak energy condition and may be unstable, so
that the phantom universe may evolve to the Big Rip singularity
\cite{Caldwell,McInnes}. In the scalar phantom model with
$\mathcal{L}=-\frac{1}{2}\dot{\phi}^2-V(\phi)$, for a constant
equation of state $w_X$, there are the scale factor $a\sim
(t_{br}-t)^{2/3(1+w_X)}$ and the field $\phi\sim ln(t-t_{br})$
with $t_{br}$ a Big Rip time \cite{Caldwell,McInnes}. By an
appropriate choice of the equation of state $w_X$, there can
always be $a(t_{br}-t)=a(t-t_{br})$, but the field isn't
well-defined for $t>t_{br}$. Clearly, the above phantom solution
$(a,\phi)$ only describes the phantom universe before the Big Rip.
For it, we can have the two possible conclusions: Either such a
solution $(a,\phi)$ describes an actual Big Rip which indicates a
final state of the universe, or it should be incomplete since the
field can not well-defined for $t>t_{br}$ providing that the Big
Rip isn't a final state.

What a Big Rip really means is still puzzling. Taking into account
the effects of the quantum gravity, the correction to the behavior
of phantom can become important near the singularity
\cite{Nojiri,SNojiri,Odintsov}, so that the Big Rip of the phantom
universe may be evaded or moderates at least. The latter case
implies that the quantum effects should only give a suppression on
the Big Rip derived in a classical gravitational theorem. In other
words, a Big Rip may actually indicate a critical state of phantom
universe between expansion and extract phase if a Big Rip means an
epoch that the quantum effects dominate, or it should still be the
final state of the universe providing that the quantum effects
only is subordinate and can not stop the phantom universe evolving
to the future sudden singularity.

The $SO(1,1)$ model shows some distinctive features \cite{Wei}.
For the exponential potential, the phantom universe may evade the
future sudden singularity and eventually settle into the de Sitter
phase \cite{WYH}. In this paper, we will find a new class of the
Big Rip in the $SO(1,1)$ phantom model, for which the scale factor
blows up but the field is regular at the Big Rip. In Sec. II, we
analyze the late-time behaviors of the scale factor and the field
for the inverse linear potential by using some approximation
conditions, and derive the late-time field, the late-time scale
factor and the late-time energy densities. In Sec. III, we first
show the late-time phantom equation of state, and then give some
discussions of the Big Rip. Here, the Big Rip may be considered as
a final state or treated as a critical point from expansion to
contract phase. The former situation is conjectured to describe a
pair of the universes having the common Big Rip.

\section{Late-time universe with inverse power-law potential}

The dark energy model with the Lagrangian
$\mathcal{L}=\frac{1}{2}(\dot{\Phi}^2-\dot{\theta}^2\Phi^2)-V$ is
derived from the $SO(1,1)$ model by defining
$\Phi=\sqrt{\phi_1^2-\phi_2^2}$ and
$\tanh\Theta=\frac{\phi_2}{\phi_1}$, where $\phi_1$ and $\phi_2$
are the two components of the $SO(1,1)$ field, $\Phi$ and $\theta$
can be called the norm and the rotation angle of the $SO(1,1)$
vector (field). The $SO(1,1)$ model may also be considered as a
generalization to the quintessence \cite{WYH,Wei}, in other words,
the quintessence corresponds to the case of the constant $\theta$
in the $SO(1,1)$ model. In this model, the field behaves like
phantom for $\dot{\Phi}^2<\dot{\theta}^2\Phi^2$ and quintessence
for $\dot{\Phi}^2>\dot{\theta}^2\Phi^2$. Here, we will focus on
the phantom case and discuss the late-time behavior of the field
which is sufficient to analyze the problem of the Big Rip. For a
spatially flat, isotropic and homogeneous universe, the late time
Einstein equation and equations of motion for $\Phi$ and $\theta$
read \cite{Wei}
\begin{eqnarray}
H^2=(\frac{\dot{a}}{a})^2=\frac{8\pi G}{3}\rho_\Phi, \label{eq1}
\end{eqnarray}
\begin{eqnarray}
\ddot{\Phi}+3H\dot{\Phi}+\dot{\theta}^2\Phi+V'(\Phi)=0,
\label{eq2}
\end{eqnarray}
\begin{eqnarray}
\dot{\theta}=\frac{c}{a^3\Phi^2}, \label{eq3}
\end{eqnarray}
and the energy density and pressure are given by
\begin{eqnarray}
\rho_\Phi=\rho_k+\rho_c+V, \quad p_\Phi=\rho_k+\rho_c-V,
\label{eq4}
\end{eqnarray}
where
\begin{eqnarray}
\rho_k=\frac12\dot{\Phi}^2, \quad
\rho_c=-\frac12\Phi^{2}\dot{\theta}^{2}, \label{eq5}
\end{eqnarray}
a dot and a prime denote $\frac{\partial}{\partial t}$ and
$\frac{\partial}{\partial \Phi}$, respectively. The constant $c$
resembles the $U(1)$ charge $Q$ in the spintessence model
\cite{Boyle}, which is the $SO(1,1)$ charge and we henceforth mark
as $\bar{Q}$.

From Eq. (\ref{eq4}), one can see
$w_\Phi=\frac{p_\Phi}{\rho_\Phi}<-1$ for $\rho_k+\rho_c<0$ and
thus for this case the model behaves like phantom. For the
exponential potential the late-time universe appears to be stable
and will finally be in the de Sitter expansion phase \cite{WYH}.
Another potential studied greatly in quintessence, phantom,
tachyon and other models \cite{Rathra,Frieman,Hao,Feinstein,Guo}
is the inverse power potential. Here, we will consider the
following potential of the form
\begin{eqnarray}
V=V(\Phi)=V_0/\Phi, \label{eq6}
\end{eqnarray}
with the constant $V_0>0$ having the unit $M^{5}$ and $M$ energy
unit.

Eq. (\ref{eq2}) has an extra term $\dot{\theta}^2\Phi$ in contrast
to the equation of motion of quintessence, which proves to play a
crucial role for the phantom case. For convenience, introduce
$\eta_1$ by
\begin{eqnarray}
\ddot{\Phi}+3H\dot{\Phi}=\eta_1\dot{\theta}^2\Phi, \label{eq7}
\end{eqnarray}
where $\eta_1$ is assumed to be a small quantity for late time.
Putting Eq. (\ref{eq7}) and $V'=-V_0\Phi^{-2}$ in Eq. (\ref{eq2})
leads to the following equation
\begin{eqnarray}
(1+\eta_1)\bar{Q}^2a^{-6}-V_0\Phi=0, \label{eq8}
\end{eqnarray}
which yields the scale factor
\begin{eqnarray}
a=[\frac{V_0\Phi}{\bar{Q}^2(1+\eta_1)}]^{-\frac{1}{6}}.
\label{eq9}
\end{eqnarray}
From (\ref{eq9}), it follows the Hubble parameter
\begin{eqnarray}
H=\frac{1}{6}[-\dot{\Phi}\Phi^{-1}+\dot{\eta}_1(1+\eta_1)^{-1}],
\label{Hubble}
\end{eqnarray}
the first term on the left-hand of which takes a dominant place
for late time, as will be seen.

Defining another parameter $\eta_2$ by
\begin{eqnarray}
\eta_2=-\frac{\rho_k}{\rho_c}, \label{eq11}
\end{eqnarray}
with $\rho_c=-\frac{V}{2(1+\eta_1)}$, which is also assumed to be
a small quantity for late time, and substituting the total energy
density $\rho_\Phi=\rho_k+\rho_c+V$ in Eq.(\ref{eq1}), then we
have
\begin{eqnarray}
H=\pm\mu_P^{-1}\delta\sqrt{V}, \label{eq12}
\end{eqnarray}
with
$\delta=\sqrt{1-\frac{1-\eta_2}{2(1+\eta_1)}}\simeq\frac{1}{\sqrt{2}}[1+\frac{1}{2}(\eta_1+\eta_2)]$
for late time, $\mu_P=\sqrt{3}M_P$ and $M_P=1/\sqrt{8\pi G}$ the
reduced Planck mass, where the sign $+$ and $-$ correspond to the
cases of expansion and contract universe, respectively. Combining
Eqs. (\ref{Hubble}) and (\ref{eq12}) yields the following equation
\begin{eqnarray}
-\frac{1}{6}[\dot{\Phi}\Phi^{-1}+\xi]=\pm \delta
H_P\Phi^{-\frac{1}{2}}, \label{eq13}
\end{eqnarray}
with $H_P=\frac{\sqrt{V_0}}{\mu_P}$, where
$\xi=\dot{\eta}_1(1+\eta_1)^{-1}$, which takes the form
$\xi\simeq\dot{\eta}_1(1-\eta_1)$, and $\delta$ can vary very
slowly for late time, noticing that $\eta_1$ and $\eta_2$ have
been assumed to be two small quantities for late time.

It is difficult to find exact solution of (\ref{eq13}). The direct
and useful way to determine whether the universe has a future
singularity is to analyze the late-time behavior of the solution.
In the following, we will only discuss the solution of
(\ref{eq13}) in the late-time situation. Treating $\delta$ as a
constant, then from Eq. (\ref{eq13}) we obtain
\begin{eqnarray}
\Phi=[\mp3\delta H_P(t-t_{br})+\zeta]^{2}, \label{eq14}
\end{eqnarray}
where $t_{br}$ is an integration constant and has the time unit,
$\zeta$ is given by
\begin{eqnarray}
\zeta=-\frac{1}{2}\int\xi\Phi^{\frac{1}{2}}dt. \label{eq15}
\end{eqnarray}
Substituting (\ref{eq14}) in Eq. (\ref{eq9}), we obtain the
late-time scale factor
\begin{eqnarray}
a=[\frac{V_0}{\bar{Q}^2(1+\eta_1)}]^{-\frac{1}{6}} [\mp3\delta
H_P(t-t_{br})+\zeta]^{-\frac{1}{3}}. \label{eq16}
\end{eqnarray}
Noting that $\delta\simeq\frac{1}{\sqrt{2}}$ for late time and
neglecting $\zeta$, then (\ref{eq14}) and (\ref{eq16}) reduce to
\begin{eqnarray}
\Phi\simeq\frac{3V_0}{2M_P^2}(t-t_{br})^{2}, \quad a\simeq
(\frac{\sqrt{3}V_0}{\sqrt{2}M_P})^{-\frac{1}{3}}[\mp\bar{Q}^{-1}(t-t_{br})]^{-\frac{1}{3}},
\label{eqfs}
\end{eqnarray}
where the charge $\bar{Q}$ is assumed to be positive, the sign $-$
is taken for $t<t_{br}$ and $+$ for $t>t_{br}$, which imply the
Hubble parameter $H>0$ for $t<t_{br}$ and $H<0$ for $t>t_{br}$.
The variations of the late-time scale factor and field with time
are shown in Fig.1.

\begin {figure}
\begin{center}
\includegraphics{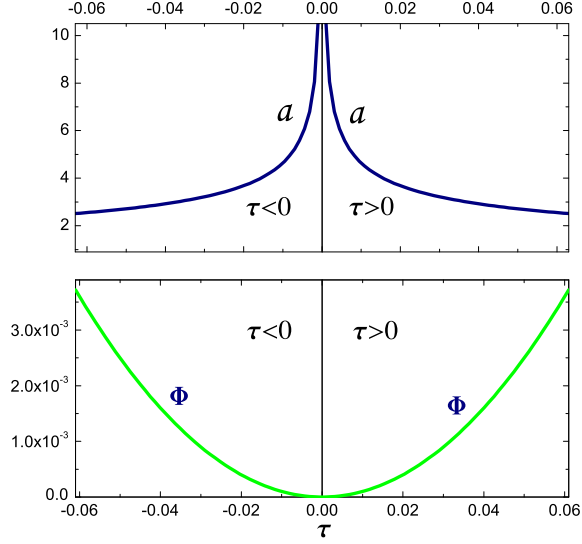}
\caption{In this figure, $\tau$ is defined as $\tau=t-t_{br}$, the
scale factor $a$ and the field $\Phi$ are given in the units
$\frac{3V_0}{2M_P^2}$ and
$(\frac{\sqrt{3}V_0}{\sqrt{2}\bar{Q}M_P})^{-\frac{1}{3}}$,
respectively.} \label{Fig.1}
\end{center}
\end{figure}

The approximation expression (\ref{eqfs}) for $\Phi$ and $a$ is
sufficient for deriving the leading terms of the late-time
$\rho_k$, $\rho_c$ and $V$. Putting $a$ and $\Phi$ given in
(\ref{eqfs}) into Eqs. (\ref{eq5}) and (\ref{eq6}), then we obtain
\begin{eqnarray}
\rho_k\simeq\frac{9}{2}V_0^2M_P^{-4}(t-t_{br})^{2}, \label{eq18}
\end{eqnarray}
\begin{eqnarray}
\rho_c\simeq -\frac{1}{3}M_P^2(t-t_{br})^{-2}, \label{eq19}
\end{eqnarray}
\begin{eqnarray}
V\simeq \frac{2}{3}M_P^2(t-t_{br})^{-2}, \label{eq20}
\end{eqnarray}
which show that $\rho_c$ and $V$ become infinite and $\rho_k$
diminishes to zero as $t\rightarrow t_{br}$, and give the
late-time energy density of the universe
$\rho_\Phi=\rho_k+\rho_c+V\simeq\frac{1}{3}M_P^2(t-t_{br})^{-2}$.
Noting that $\dot{\theta}^2\Phi=-2\rho_c\Phi^{-1}$ and the
late-time Hubble parameter $H\simeq -\frac{1}{3}(t-t_{br})^{-1}$,
then from (\ref{eqfs}) there are
\begin{eqnarray}
\dot{\theta}^2\Phi=\frac{4}{9}V_0^{-1}M_P^{4}(t-t_{br})^{-4},
\quad \ddot{\Phi}=3V_0M_P^{-2}, \quad 3H\dot{\Phi}=-3V_0M_P^{-2}.
\label{eq21}
\end{eqnarray}

Clearly,
$\eta_1=\frac{\ddot{\Phi}+3H\dot{\Phi}}{\dot{\theta}^2\Phi}\simeq0$
and $\eta_2=-\frac{\rho_k}{\rho_c}\sim O[(t-t_{br})^{4}]$ are
indeed the two small quantities for late time, as anticipated
above. From $\xi\simeq \dot{\eta}_1(1-\eta_1)$ and $\Phi$ in
(\ref{eqfs}), it follows that $\zeta\sim 0$ for late time. Thus,
the scale factor $a$ and the field $\Phi$ in Eq. (\ref{eqfs}) and
the energy densities $\rho_k$, $\rho_c$ and $V$ given in Eqs.
(\ref{eq18})-(\ref{eq20}) become more and more precise as time
approaches the Big Rip.

\section{Conclusions and discussions}

In the previous section, it has been shown that the late-time
$\rho_c$ and $V$ are quadratically divergent and $\rho_k$
approaches zero as $t\rightarrow t_{br}$. From Eqs.
(\ref{eq18})-(\ref{eq20}), the late-time equation of state is
given by
$w_\Phi=\frac{\rho_k+\rho_c-V}{\rho_k+\rho_c+V}\simeq-3+54V_0^2M_P^{-6}(t-t_{br})^{4}$,
which reaches its minimum $-3$ at the Big Rip. Eqs.
(\ref{eq18})-(\ref{eq20}) show that near the Big Rip the total
energy density of the universe grows according to the law
$(t-t_{br})^{-2}$ same as that in given in
\cite{Caldwell,McInnes}. Eq. (\ref{eqfs}) also appears the
late-time scale factor obeys the same law $a\sim
(t-t_{br})^{2/3(1+w_\Phi)}$ given in terms of $w_\Phi$ for a
constant equation of state. However, the scale factor in
(\ref{eqfs}) can contain an additional information from the charge
$\bar{Q}$, as will seen in the following. Besides, the
descriptions for the dynamics fields are also quite different, in
the scale phantom model the field has a logarithmatic divergence
at the Big Rip, while in the $SO(1,1)$ phantom model the field can
be regular at it, as shown in Fig.1.

The Big Rip may imply the nonanalyticity of the scale factor
resulted from the model itself, which may be removed in the
quantum gravity theorem \cite{Odintsov}. If so, then the Big Rip
can be specified as a critical state of the universe. For this
treatment, a phantom universe will undergo the two different
stages: the expansion and the contract phase. For us, such a Big
Rip should be viewed as the end of our universe, but it can look
like an initial state for the living who uses $\tau=0$, instead of
$t=0$, as the beginning of the time. Alternatively, the Big Rip
may be considered as the final state \cite{Caldwell,McInnes}. For
$t>t_{br}$, from Eq. (\ref{eqfs}), we have
\begin{eqnarray}
\Phi\simeq\frac{3V_0}{2M_P^2}\tau^{2}, \quad a\simeq
(\frac{\sqrt{3}V_0}{\sqrt{2}M_P})^{-\frac{1}{3}}(\bar{Q}^{-1}\tau)^{-\frac{1}{3}},
\label{eq22}
\end{eqnarray}
with $\tau=t-t_{br}$. Eq. (\ref{eq22}) shows clearly the symmetry
transformation $\tau\rightarrow-\tau$ (the time-reversal
transformation for the redefined time $\tau$, which is still
called $T$, here) and $\bar{Q}\rightarrow-\bar{Q}$ (looking like
charge conjugate transformation in particle physics, which is also
called $C$) about the Big Rip. As a result, the field is
$T$-invariant and the scale factor is $CT$-invariant. This
symmetry can induce us to imagine the Big Rip as an infinite plane
mirror, to which the scale factor has the imaginary image. This
can suggest a pair of the universes characterized by the charge
$\bar{Q}$ and having the common Big Rip, one with $\bar{Q}>0$ is
the universe on which we live and for us the other with
$\bar{Q}<0$ is the image universe which may be an actual universe
for the human being who live there.

In summary, there evidently are some inviting characteristics for
the $SO(1,1)$ Big Rip. As a mathematic interest, the field and the
scale factor are well-defined both in the range $t>t_{br}$ and
$t<t_{br}$; The field becomes zero at the Big Rip, instead of
infinity, this indicates a fundamental difference from the one
given in the single field phantom model for which both the scale
factor and the field blow up; The field possesses $T$ symmetry and
the scale factor has $CT$ symmetry, which can suggest a pair of
the universes having the same final state.

\vskip 1.0cm

{\bf Acknowledgement:} This work is supported by Liaoning Province
Educational Committee Research Project (20040026) and National
Nature Science Foundation of China; it was supported ITP
Post-Doctor Project (22B580), Chinese Academy of Science of China.

\end{document}